\documentclass[%
 reprint,
 amsmath,amssymb,
 aps,
]{revtex4-2}

\usepackage{graphicx}
\usepackage{dcolumn}
\usepackage{bm}
\usepackage{color,soul,url}

\begin{document}

\preprint{APS/123-QED}

\title{The Chern Number Governs Soliton Motion in Nonlinear Thouless Pumps}

\author{Marius J\"urgensen}
 \email{marius@psu.edu}
\author{Mikael C. Rechtsman}%
 \email{mcrworld@psu.edu}
\affiliation{%
Department of Physics, The Pennsylvania State University, University Park, Pennsylvania 16802, USA}%

\date{\today}

\begin{abstract}
Nonlinear Thouless pumps for bosons exhibit quantized pumping via soliton motion, despite the lack of a meaningful notion of filled bands. However, the theoretical underpinning of this quantization, as well as its relationship to the Chern number, has thus far been lacking. Here we show that for low power solitons, transport is dictated by the Chern number of the band from which the soliton bifurcates. We do this by expanding the discrete nonlinear Schr\"odinger equation (equivalently, the Gross-Pitaevskii equation) in the basis of Wannier states, showing that the soliton’s position is dictated by that of the Wannier state throughout the pump cycle. Furthermore, we describe soliton pumping in two dimensions.

\end{abstract}

\maketitle
The concept of Thouless pumping \cite{Thouless1983a,Niu1984,Xiao2010,Asboth2016} has played a major role in shaping the understanding of the robustness of the integer quantum Hall effect and its salient feature: Sharply quantized plateaus in the Hall conductance despite the presence of disorder. After it was first observed experimentally \cite{Klitzing1980}, a full theoretical description was much sought after. Laughlin put forward a magnetic-flux threading argument, in which charge is pumped from one edge of a two-dimensional system to the other due to gauge invariance \cite{Laughlin1981}. Thouless gave an alternative theoretical explanation via dimensional reduction (as it was termed later) and showed that the quantized conductance in two dimensions maps onto quantized pumping in a 1+1-dimensional system, which is periodically modulated in time \cite{Thouless1983a,Niu1984}. In such `Thouless pumps', quantized pumping of charge can be understood in terms of Wannier functions that are displaced by an amount dictated by a topological invariant of the system \cite{Zak1989,Asboth2016}: the Chern number of the corresponding band \cite{Thouless1982,Thouless1983a,Simon1983,Niu1984}.

More recently, it came to be understood that topological states such as those associated with the quantum Hall effect are not restricted to fermions, but are rather a general wave phenomenon \cite{Raghu2008,Haldane2008a}. Besides the hallmark electronic systems \cite{Klitzing1980,Tsui1982}, topological phenomena have been predicted and observed in a variety of experimental platforms based on fermions and bosons, such as ultracold atoms \cite{Atala2013,Jotzu2014,Aidelsburger2015}, mechanical systems \cite{Nash2015,Suesstrunk2015} and photonics \cite{Wang2009,Rechtsman2013,Hafezi2013}. Moreover, linear Thouless pumps have been observed in such systems (see e.g., Refs. \cite{Kraus2012,Verbin2013, Lohse2016,Nakajima2016,Lohse2018,Zilberberg2018,Grinberg2020} to cite a few). An important area of contemporary research is the interplay between topology and inter-particle interaction; this has been highly challenging due to the lack of a broad theoretical framework for topological invariants and associated physical observables in such systems. Interacting bosons in the mean-field limit (whether photons, atoms, or otherwise) are usually described by a nonlinear Schr\"odinger equation -- also called Gross-Pitaevskii equation -- where the nonlinear term is related to the strength of the inter-particle interactions. Perhaps the most fascinating solution of nonlinear equations are  solitons \cite{Askaryan1962, Chiao1964, Ablowitz1981, Christodoulides1988, Stegeman1999}. These are wavepackets localized in space and eigenstates of their self-induced potential. In other words, the nonlinearity acts to balance the diffraction and nonlinearity, such that a soliton's shape is constant in time or space. The interplay between nonlinearity/solitons and topology has recently led to discoveries of new phenomena such as Floquet topological bulk solitons \cite{Lumer2013,Mukherjee2020a}, soliton-like edge propagation \cite{Ablowitz2014,Leykam2016,Mukherjee2020b} and nonlinearly-induced edge states \cite{Maczewsky2020}. Recently, the authors observed quantized nonlinear Thouless pumping via soliton motion despite non-uniform band occupation \cite{Juergensen2021}. The quantization followed from the fact that the Hamiltonian was time-periodic and thus came back to itself after a period, implying that the soliton wavefunction must -- in the low-power regime -- return to itself (apart from a translation by an integer number of unit cells). In that work, it was observed experimentally and numerically that solitons bifurcating from a given band are transported in accordance with the Chern number of that band. However, a comprehensive theoretical understanding of this effect has not yet been presented.

\begin{figure}[b]
\includegraphics{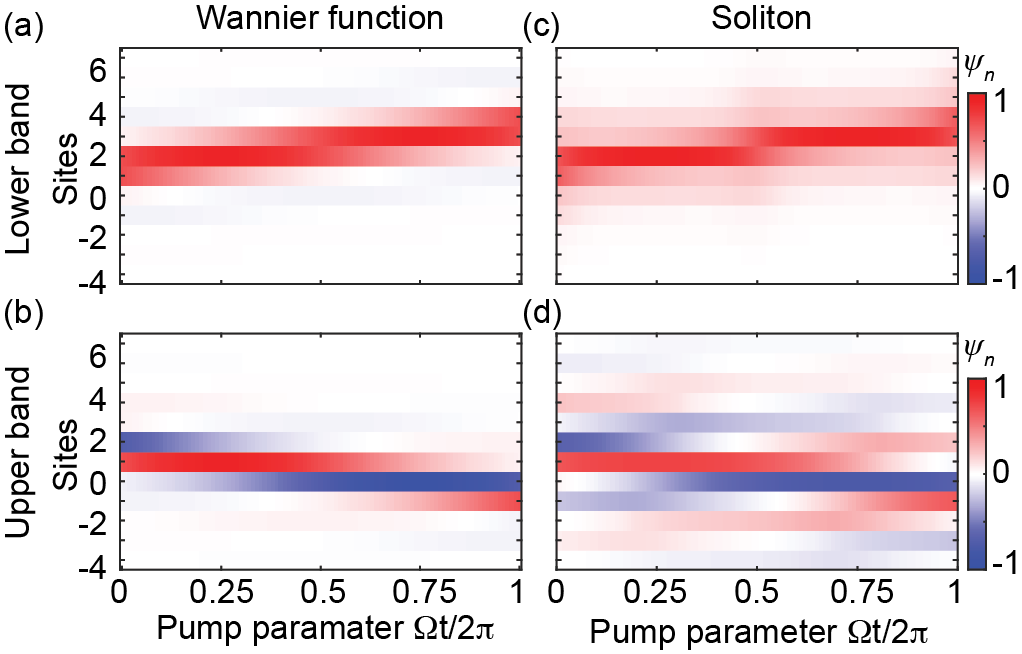}
\caption{\label{fig1} Wannier states and soliton wavefunctions. (a) Instantaneous maximally localized Wannier wavefunction for one pump cycle calculated for the lower band in a Rice-Mele model with Chern number $C=+1$. (b) Similar to (a), but calculated for the upper band with a Chern number of $C=-1$. (c,d) Similar to (a,b) but showing the instantaneous soliton wavefunction. The degree of nonlinearity is $g/J=2$ and $g/J=3$ (see Eq. \ref{Eq:Rice-Mele}) for (c) and (d), respectively.}
\end{figure}

Here we show that the trajectory of low-power solitons in nonlinear Thouless pumps is dictated by the Chern number of the band from which they bifurcate. Because for adiabatic modulation a stable soliton remains a soliton of the instantaneous Hamiltonian, it is sufficient to show the existence of a stable soliton for all times, and that its position is linked to the position of the Wannier states. Specifically, we solve for the instantaneous solitons in the basis of Wannier functions, in which the equations take the form of a simple 1D lattice with Kerr nonlinearity. We illustrate the resulting quantization in a Rice-Mele model. For each band we find a stable instantaneous soliton centered upon a Wannier function for all times during the pump cycle. Additionally, we find an unstable soliton centered between two Wannier functions. Finally, we show quantized nonlinear pumping in a 2+2-dimensional pump.

Our focus lies on systems with a slowly varying, time-periodic Hamiltonian describing a (linear) Thouless pump. The pump is entirely general at this stage; later we illustrate the results in a Rice-Mele model. The time-dynamics are described by the discrete nonlinear Schr\"odinger equation with a focusing Kerr nonlinearity \cite{Kivshar2003,Kevrekidis2009} (although the results generalize straightforwardly to the defocusing case):
\begin{equation}
i \dot{\psi}_n(t) = \sum_m H_{nm}(t) \psi_m(t) - g |\psi_n(t)|^2 \psi_n(t)
\label{DNLSE_Sites}
\end{equation}
Here, $\psi_n(t)$ is the amplitude of the wavefunction at site $n$ and time $t$, $H_{nm}(t)$ is a time-periodic tight-binding Hamiltonian, $g>0$ is the strength of the focusing nonlinearity and $\dot{\psi}_n$ represents the time derivative of $\psi_n$. For our analysis it does not make a difference if $H_{nm}$ describes hoppings between sites or orbitals on a given site. Thus, we refer to both as sites. The indices $n$, $m$ run over all sites with periodic boundary conditions. Eq. (\ref{DNLSE_Sites}) describes a range of systems, including the propagation of intense light through nonlinear media \cite{Kevrekidis2009}, the dynamics of Bose-Einstein condensates \cite{Dalfavo1999} and exciton-polariton condensates \cite{deng2002, kasprzak2006, balili2007}. It conserves the norm of the wavefunction ($P=\sum_n{|\psi_n|^2}$) and without loss of generality, we use normalized wavefunctions ($P=1$) and vary $g$. The degree of nonlinearity is then given by $g$ with respect to the hopping parameters in $H_{nm}$. We refer to solitons as having `low power' if they are calculated with small $g$.

In the linear case ($g=0$) quantized Thouless pumping necessarily requires adiabatic time modulation and uniform band occupation. For fermionic systems (like electrons in the solid state), this corresponds to a Fermi level within a band gap. For bosonic systems, a simple way to obtain uniform band filling is via the initial excitation of a single Wannier function. Over time, the wavefunction will evolve but retain a uniform band occupation throughout the pump cycle as dictated by the adiabatic theorem. Quantized pumping itself can be understood as flow of the instantaneous Wannier functions as displayed in Figs. 1a,b, whose winding around the unit cell as a function of the pump parameter is equivalent to the Chern number of the occupied band.

In the nonlinear case ($g \neq 0$) the time evolution of stable solitons (i.e., nonlinear eigenstates) has similarities to the adiabatic time evolution of eigenstates in linear systems \cite{Liu2003,Pu2007}: For sufficiently slow driving, excitations of other states are negligible and the wavefunction continues to occupy the instantaneous soliton for all times during the pump cycle. It is therefore possible to calculate the adiabatic time-evolution for solitons in two ways: (1) Numerically solving Eq. (\ref{DNLSE_Sites}) as a function of time; or (2) solving for the instantaneous nonlinear eigenstates at different time slices in the pump cycle. Importantly, the instantaneous states must be stable, as otherwise small perturbations around the linearized solution exponentially increase (see for example \cite{Kevrekidis2009} for linear stability analysis). We illustrate the second method in Figs. 1c,d, which show the wavefunction of instantaneous solitons. Strikingly, the trajectories of the instantaneous solitons are noticeably similar to those of the instantaneous Wannier functions of the bands from which the solitons bifurcate. As the available solitons at the beginning and at the end of each pump cycle are identical in each unit cell, quantized motion of solitons is expected, even for non-uniform band occupation. Below, we prove that the number of unit cells which the solitons are pumped corresponds to the Chern number of the band from which they bifurcate.

We now present the missing link to show that the position of the instantaneous soliton is indeed intimately related to that of the Wannier functions.  Showing that the solitons pump by the same number of unit cells as the Wannier functions will prove that the solitons are transported by $C$ unit cells, where $C$ is the Chern number. As we are only concerned with finding the instantaneous solitons for a static Hamiltonian at a given point in the pump cycle, we use $\psi_n(t) \rightarrow e^{-i\lambda t} \psi_n$, where $\lambda$ is the nonlinear eigenvalue, such that Eq. (1) takes the following form:
\begin{equation}
\lambda \psi_n = \sum_m H_{nm} \psi_m - g |\psi_n|^2 \psi_n
\label{DNLSE_SitesStatic}
\end{equation}

We first rewrite Eq. (\ref{DNLSE_SitesStatic}) in the basis of Wannier functions as in Ref. \cite{Alfimov2002} by expanding the wavefunction in the Wannier basis:
\begin{equation}
    \psi_n = \sum_{\textbf{R},\alpha} c_{\textbf{R},\alpha} w_{\textbf{R},\alpha,n}
\label{expansionInWannier}
\end{equation}
with expansion coefficients $c_{\textbf{R},\alpha}$ and Wannier functions $w_{\textbf{R},\alpha,n}$, labelled by the lattice vector $\textbf{R}$ and a band index $\alpha$. By plugging Eq. (\ref{expansionInWannier}) into Eq. (\ref{DNLSE_SitesStatic}) and after some mathematical manipulation we arrive at the following equation (for a more detailed derivation, see the Supplemental Material):
\begin{multline}
    \lambda c_{\textbf{R},\alpha} = \sum_{\textbf{R}^{\prime}} \epsilon_{\textbf{R}^{\prime}+\textbf{R},\alpha} c_{\textbf{R}^{\prime},\alpha}+ \\
    g W_{\textbf{R},\textbf{R}^{\prime},\textbf{R}^{\prime\prime},\textbf{R}^{\prime\prime\prime}}^{\alpha,\alpha^{\prime},\alpha^{\prime\prime},\alpha^{\prime\prime\prime}} c_{\textbf{R},\alpha}^* c_{\textbf{R}^{\prime\prime},\alpha^{\prime\prime}}c_{\textbf{R}^{\prime\prime\prime},\alpha^{\prime\prime\prime}}
\label{DNLSE_WanFull}
\end{multline}
Here, $\epsilon_{\textbf{R},\alpha}$, is the Fourier coefficient of the energy of band $\alpha$, and $W$ is an overlap integral between four Wannier functions:
\begin{equation}
    W_{\textbf{R},\textbf{R}^{\prime},\textbf{R}^{\prime\prime},\textbf{R}^{\prime\prime\prime}}^{\alpha,\alpha^{\prime},\alpha^{\prime\prime},\alpha^{\prime\prime\prime}} = \sum_n w_{\textbf{R},\alpha,n}^* w_{\textbf{R}^{\prime},\alpha^{\prime},n}^* w_{\textbf{R}^{\prime\prime},\alpha^{\prime\prime},n} w_{\textbf{R}^{\prime\prime\prime},\alpha^{\prime\prime\prime},n}
\end{equation}

Up to now, no assumptions have been made and Eq. (\ref{DNLSE_WanFull}) and Eq. (\ref{DNLSE_SitesStatic}) are equally valid to find static solitons, apart from the fact that they are written in different bases: $\psi_n$ describes the amplitude on individual real space sites, while $c_{\textbf{R},\alpha}$ describes the amplitude of Wannier functions. We refer to the two descriptions as `real space' and `Wannier space', respectively. To proceed, we make the following reasonable simplifications:
(1) We focus on static solitons at low power, whose occupation for isolated (non-degenerate) energy bands tends to be in one band only, due to the large energy difference separating the bands. Therefore, we can neglect any nonlinear inter-band coupling terms in Eq. (\ref{DNLSE_WanFull}) and only focus on individual bands.
(2) We restrict ourselves to systems that allow exponentially localized Wannier functions, such that the dominant term in the overlap integral is given by the Wannier functions localized in the same unit cell, which applies to a Thouless pump at any given time slice.

Then, the static discrete nonlinear Schr\"odinger equation takes the following fully simplified form in the Wannier basis:
\begin{equation}
    \lambda c_{\textbf{R},\alpha} = \sum_{\textbf{R}^{\prime}} \epsilon_{\textbf{R}^{\prime}-\textbf{R},\alpha} c_{\textbf{R}^{\prime},\alpha} + g W_{\textbf{R},\textbf{R},\textbf{R},\textbf{R}}^{\alpha,\alpha,\alpha,\alpha} |c_{\textbf{R},\alpha}|^2 c_{\textbf{R},\alpha}
\label{DNLSE_WanSimple}
\end{equation}
This equation has an intuitive interpretation: It describes hopping in a lattice with a Kerr nonlinearity, where the hopping strengths are given by the Fourier coefficients of the (linear) energy bands. In contrast to Eq. (\ref{DNLSE_SitesStatic}), the sites in this lattice do not represent sites in real space, but rather Wannier functions. Furthermore, and importantly, the unit cell in the Wannier space only consists of a single site. This allows us to use the knowledge about solitons in simple 1D lattices (see for example Refs. \cite{Christodoulides1988, Kivshar1993,Kivshar2003,Kevrekidis2009}) where we already know that two types of static solitons exist: One stable on-site soliton and one unstable inter-site soliton (see Fig. 2a). In Wannier space the on-site soliton is always centered on a single Wannier function. After transforming back from Wannier into real space, the obtained real space soliton is localized near the Wannier center. (Its center of mass is not necessarily identical to the center of mass of the Wannier function due to interference effects between multiple occupied Wannier functions.) As this is valid for all times $t$ during a pump cycle, and the on-site soliton is expected to be stable for all times, we have shown that the motion of the soliton is dictated by the motion of the Wannier centers. Thus, the displacement of low-power pumped solitons is therefore given by the Chern number of the band from which they bifurcate.

\begin{figure}[b]
\includegraphics{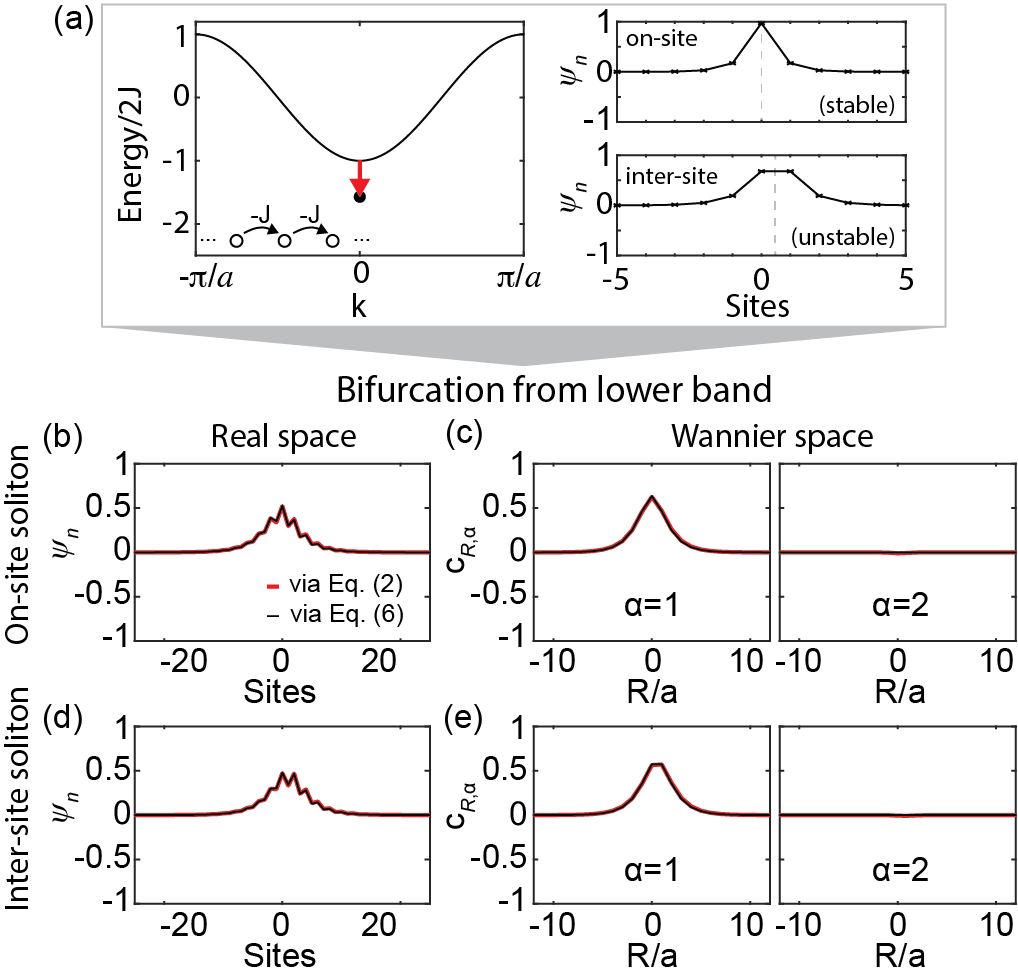}
\caption{\label{fig2} Comparison between soliton calculation in real space via Eq. (\ref{DNLSE_SitesStatic}) and Wannier space via Eq. (\ref{DNLSE_WanSimple}). (a) Left panel: Band structure for a simple 1D lattice with hopping $J>0$. Bifurcation of solitons (for focusing nonlinearity) occurs from the minimum of the band. Right panels: The two types of possible solitons: on-site and inter-site soliton. (b,c) Expansion coefficients, $\psi_n$ and $c_{\textbf{R},\alpha}$, of the instantaneous on-site soliton in real space (b) and Wannier space (c) for $\Omega t=2\pi/8$ and the lower band in a Rice-Mele model (see Eq. (\ref{Eq:Rice-Mele})) with $g/J=1$. The soliton calculated in real space via Eq. (\ref{DNLSE_SitesStatic}) is shown in red. The soliton calculated in Wannier space via Eq. (\ref{DNLSE_WanSimple}) is shown in black. $\alpha$ is the band index. (d,e) Similar to (b,c) but for the inter-site soliton.}
\end{figure}


To numerically illustrate our findings by example, we use a Rice-Mele model \cite{Rice1982} with a focusing nonlinearity ($g>0$): The linear Hamiltonian of this model describes a Thouless pump with two sites per unit cell:
\begin{align}
\begin{split}
    H_{nm}^{\text{RM}}(t) = -&[J + (-1)^{m+1} {} \delta \cos(\Omega t)] \delta_{n-1,m}\\
    -& [J + (-1)^{m} \delta \cos(\Omega t)] \delta_{n+1,m}\\
    -& {} \Delta (-1)^{m} \sin(\Omega t) \delta_{n,m}
\end{split}
\label{Eq:Rice-Mele}
\end{align}
Here, $\Omega$ is the modulation frequency. The parameter $J$ describes the average hopping strength between nearest-neighbor sites, which is modulated with strength $\delta$, introducing a difference between intra- and inter-unit cell hoppings. The parameter $\Delta$ gives the strength of modulation of the staggered on-site potential. For specific numerical examples throughout this work, we use $J = 1$, $\delta = 0.5$ and $\Delta = 1$ with 200 sites and periodic boundary conditions. Further information on this Hamiltonian can be found in the Supplemental Material.

We illustrate the intimate link between the position of the soliton and the Wannier function in the Rice-Mele model for $\Omega t = 2 \pi / 8$, where no spatial symmetries pin the soliton to a fixed position.  We calculate the instantaneous solitons in two ways and then compare their shape in real and Wannier space: (1) The exact soliton (Figs. 2b-e; shown in red) is calculated via Eq. (\ref{DNLSE_SitesStatic}) in real space and then transformed via change of basis into Wannier space. (2) We use Eq. (\ref{DNLSE_WanSimple}) and calculate the soliton (Figs. 2b-e; shown in black) in Wannier space, which is then transformed into real space. For low degree of nonlinearity excellent agreement is observed, both for the on-site as well as for the inter-site soliton (see Figs. 2b-e). With increasing nonlinearity, the projection of the exact soliton calculated via Eq. (\ref{DNLSE_SitesStatic}) shows more and more occupation of Wannier states of higher bands, suggesting that the approximations made in Eq. (\ref{DNLSE_WanSimple}) become less accurate (see also Supplemental Material). Solitons that bifurcate from the higher band are shown in the Supplemental Material.

\begin{figure}[b]
\includegraphics{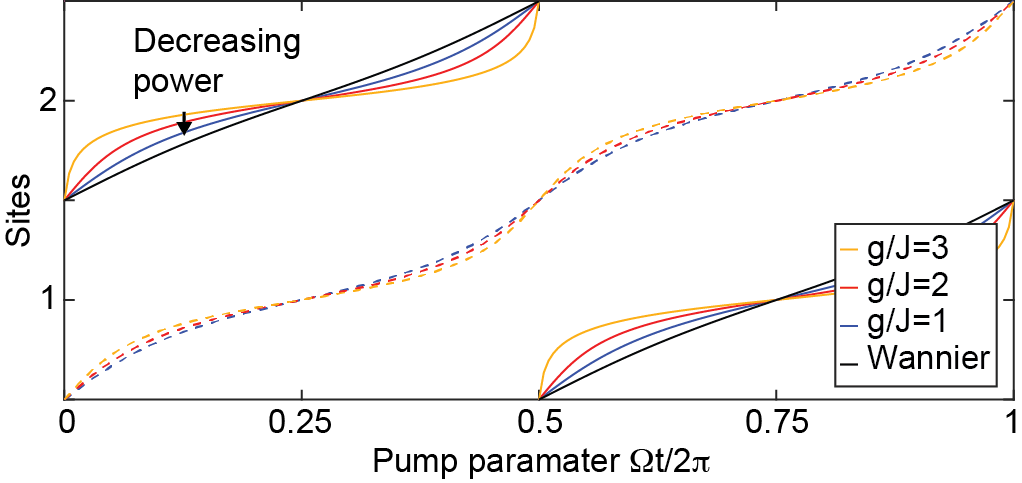}
\caption{\label{fig3} Soliton and Wannier flow. Position of the center of mass of the instantaneous solitons that bifurcate from the lower band with Chern number $C=+1$ for one pump cycle and projected into one unit-cell. Stable on-site (unstable inter-site) soliton is shown with solid (dashed) lines. Different degrees of nonlinearity are shown in color. Black line shows the center of mass position of the Wannier function.}
\end{figure}

In order to show the pumping process, we calculate the position (modulo a unit cell) of the soliton for one complete pumping cycle. Fig. 3 shows the position of the stable on-site soliton (solid lines) and the position of the unstable inter-site soliton (dashed lines) for different degrees of nonlinearity ($g/J$) calculated via Eq. (\ref{DNLSE_SitesStatic}). The stable solitons indeed follow the position of the Wannier function (black). We point out that even at low power, due to interference terms between the occupied Wannier functions, the center of mass of the soliton does not have to be centered exactly upon the center of mass of the Wannier function, but is localized close to it. For stronger nonlinearity, the approximations of Eq. (6) become less accurate and larger deviations occur, due to the increasing occupation of Wannier functions of higher bands (see also Supplemental Material). Nevertheless, the general features of the soliton are still consistent with the low-power regime. Indeed, because -- in the low power regime -- the soliton at the beginning and end of each pumping cycle is identical (apart from a translation by an integer number of unit cells), a nonlinear bifurcation is needed to either split the trajectory, or change the linear stability of the soliton (see also \cite{Juergensen2021}).


Finally, we show that quantized nonlinear pumping can be extended to higher dimensions, by using a two-dimensional model consisting of the sum of two Thouless pumps in orthogonal spatial directions. Here, we use the off-diagonal version of the Aubry-Andr\'e-Harper (AAH) model \cite{Aubry1980,Harper1955,Ke2016}. A schematic of the model is depicted in Fig. 4a where only the hoppings are modulated: $K_j = -K - \tilde{K} \cos(4\pi j/3+\Omega t)$ with $j \in \{1,2,3\}$. This model has been used to simulate the 4D quantum Hall effect and its topological properties are described by the second Chern number, which is the product of first Chern numbers for the two orthogonal directions \cite{Zilberberg2018}. The band structure for the pumping cycle is shown in Fig. 4b. We focus on a soliton that bifurcates from the lowest band. We point out that while the linear model is separable in the $x$ and $y$ directions, the nonlinear model is not. Fig 4c, shows the soliton at the beginning of the pumping cycle $\Omega t = 0$, pinned to the corner of the unit cell, due to symmetries. During the pump cycle (see Fig. 4d) the soliton is pumped by +1 unit cell in the x-direction and +1 unit cell in the y-direction, corresponding to the Chern numbers of the pumps in those directions, respectively.

\begin{figure}
\includegraphics{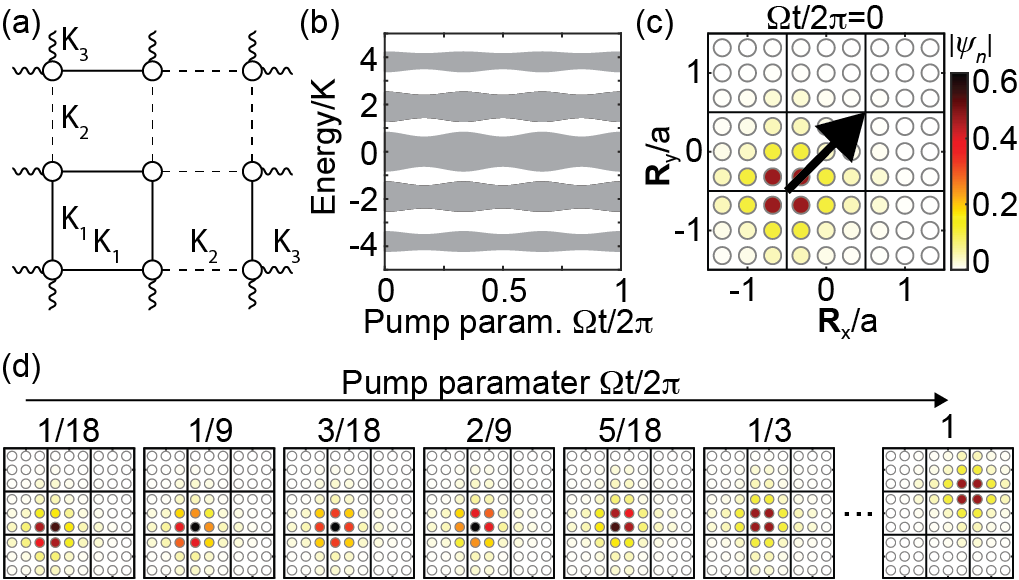}
\caption{\label{fig4} Two-dimensional quantized nonlinear pumping. (a) Schematic of the unit cell with 9 sites and three different hoppings, which are modulated in time. (b) Band structure for the system shown in (a). (c) Instantaneous soliton localized at the corner of the unit cell for $\Omega t/2\pi=0$. (d) Same as (c) but showing the movement of the soliton for evolving pump parameter. The parameters for the AAH-model are chosen as: $K=1$ and $\tilde{K} = 0.7$. In (c,d) the degree of nonlinearity is $g/K = 5$.}
\end{figure}

In summary, we have shown that solitons in weakly interacting bosonic systems are pumped by the Chern number of the band from which they bifurcate, despite non-uniform band occupation. This proves that quantized nonlinear Thouless pumping is protected by the Chern number, which can thus be considered to be a physically meaningful topological invariant for describing nonlinear systems. Soliton motion remains quantized until a nonlinear bifurcation alters the path of the soilton or makes it unstable. Furthermore, we described the Thouless pumping of unstable inter-site solitons and showed that quantized nonlinear pumping can also be observed in two-dimensional systems. Our results pave the way to a broader understanding of the interface between interacting/nonlinear systems and topology.

\vspace{15px}
During the preparation of this manuscript, the authors became aware of related, as yet unpublished work by N. Mostaan and N. Goldman.

\begin{acknowledgments}
We acknowledge fruitful discussions with S. Gopalakrishnan, N. Mostaan, N. Goldman and Panayotis Kevrekidis. We further acknowledge the support of the ONR YIP program under award number N00014-18-1-2595, ONR-MURI program N00014-20-1-2325, as well as by the Packard Foundation fellowship, number 2017-66821. Some numerical calculations were performed on the Pennsylvania State University’s Institute for Computational and Data Sciences’ Roar supercomputer.
\end{acknowledgments}


\bibliography{bibliography}

\end{document}


\preprint{APS/123-QED}

\title{Supplemental Material: The Chern Number Governs Soliton Motion in Nonlinear Thouless Pumps}

\author{Marius J\"urgensen}
 \email{marius@psu.edu}
\author{Mikael C. Rechtsman}%
 \email{mcrworld@psu.edu}
\affiliation{%
Department of Physics, The Pennsylvania State University, University Park, Pennsylvania 16802, USA}%

\date{\today}

\maketitle


\renewcommand{\thepage}{S\arabic{page}}
\renewcommand{\thesection}{S\arabic{section}}
\renewcommand{\thetable}{S\arabic{table}}
\renewcommand{\thefigure}{S\arabic{figure}}
\renewcommand{\theequation}{S\arabic{equation}}

\section{Derivation of the Discrete Nonlinear Schr\"odinger Equation in Wannier basis}

Here, we present a more detailed derivation of Eq. (4) in the main part, starting from the static discrete nonlinear Schr\"odinger equation. We restrict ourselves to systems with one spatial dimension:
\begin{equation}
    \lambda \psi_n = \sum_m H_{nm} \psi_m - g |\psi_n|^2 \psi_n
\label{S:DNLSE_SitesStatic}
\end{equation}
As described in the main text, we first expand the wavefunction in the basis of instantaneous Wannier functions:
\begin{equation}
    \psi_n = \sum_{\textbf{R},\alpha} c_{\textbf{R},\alpha} w_{\textbf{R},\alpha,n}
\label{S:WannierExpansion}
\end{equation}
with expansion coefficients $c_{\textbf{R},\alpha}$ and Wannier functions $w_{\textbf{R},\alpha,n}$ that are labelled by the lattice vector $\textbf{R}$ and a band index $\alpha$. Wannier functions form a complete and orthonormal set:
\begin{equation}
    \sum_n w_{\textbf{R}^{\prime},\alpha^{\prime},n}^* w_{\textbf{R},\alpha,n} = \delta_{\textbf{R}^{\prime},\textbf{R}} \delta_{\alpha',\alpha}
\label{S:WannierOrtho}
\end{equation}
Furthermore, the Wannier states form Fourier pairs with the Bloch states, $B_{k,\alpha,n}$, which are labelled by the crystal momentum $k$:
\begin{subequations}
\begin{equation}
    w_{\textbf{R},\alpha,n} = \frac{1}{\sqrt{2\pi}} \sum_k \exp^{-i k \textbf{R}} B_{k,\alpha,n}
\label{S:WanInReal}
\end{equation}
\begin{equation}
    B_{k,\alpha,n} = \frac{1}{\sqrt{2\pi}} \sum_{\textbf{R}} \exp^{i k \textbf{R}} w_{\textbf{R},\alpha,n}
\label{S:RealInWan}
\end{equation}
\end{subequations}
Bloch states form the energy eigenstates of the linear Hamiltonian, $H_{nm}$, with eigenenergy $E_{k,\alpha}$ and obey the following eigenvalue equation:
\begin{equation}
    E_{k,\alpha} B_{k,\alpha,n} = \sum_m H_{nm} B_{k,\alpha,m}
\label{S:EigenvalueEq}
\end{equation}
Replacing the wavefunction in Eq. (\ref{S:DNLSE_SitesStatic}) via Eq. (\ref{S:WannierExpansion}), multiplying from the left with $\omega_{\textbf{R}^{\prime},\alpha^{\prime},n}^*$, summing over all sites, and using the orthogonality relation of Wannier functions (Eq. (\ref{S:WannierOrtho})), results in:
\begin{multline}
    \lambda c_{\textbf{R}^{\prime},\alpha^{\prime}} = \sum_{\textbf{R},\alpha} c_{\textbf{R},\alpha} \sum_{n,m} \omega_{\textbf{R}^{\prime},\alpha^{\prime},n} H_{nm} \omega_{\textbf{R},\alpha,m} + \\
    g W_{\textbf{R},\textbf{R}^{\prime},\textbf{R}^{\prime\prime},\textbf{R}^{\prime\prime\prime}}^{\alpha,\alpha^{\prime},\alpha^{\prime\prime},\alpha^{\prime\prime\prime}} c_{\textbf{R},\alpha}^* c_{\textbf{R}^{\prime\prime},\alpha^{\prime\prime}}c_{\textbf{R}^{\prime\prime\prime},\alpha^{\prime\prime\prime}}
    \label{S:DNSEinWan}
\end{multline}
where $W$ is an overlap integral between four Wannier functions:
\begin{equation}
    W_{\textbf{R},\textbf{R}^{\prime},\textbf{R}^{\prime\prime},\textbf{R}^{\prime\prime\prime}}^{\alpha,\alpha^{\prime},\alpha^{\prime\prime},\alpha^{\prime\prime\prime}} = \sum_n w_{\textbf{R},\alpha,n}^* w_{\textbf{R}^{\prime},\alpha^{\prime},n}^* w_{\textbf{R}^{\prime\prime},\alpha^{\prime\prime},n} w_{\textbf{R}^{\prime\prime\prime},\alpha^{\prime\prime\prime},n}
\end{equation}
The first term on the right-hand side of Eq. (\ref{S:DNSEinWan}) can be simplified, using the Fourier series description of the band structure:
\begin{equation}
    E_{k,\alpha} = \sum_{\tilde{\textbf{R}}} e^{i k \tilde{\textbf{R}}} \epsilon_{\tilde{\textbf{R}},\alpha}
\label{S:FourierSeriesEnergy}
\end{equation}
In the Thouless pumps under consideration $E_{k,\alpha}$ is real, periodic and even around $k=0$. It follows that $\epsilon_{\tilde{\textbf{R}},\alpha} = \epsilon_{-\tilde{\textbf{R}},\alpha} = \epsilon_{\tilde{\textbf{R}},\alpha}^*$. Thus the Fourier series coefficients are a real and even function of $\tilde{\textbf{R}}$. 
With this:
\begin{flalign}
    &\sum_{\textbf{R},\alpha} c_{\textbf{R},\alpha} \sum_{n,m} \omega_{\textbf{R}^{\prime},\alpha^{\prime},n} H_{nm} \omega_{\textbf{R},\alpha,m} = \\
    &\sum_{\textbf{R},\alpha,n} c_{\textbf{R},\alpha} w_{\textbf{R}^{\prime},\alpha',n} \frac{1}{\sqrt{2\pi}} \sum_k E_{k,\alpha} e^{-i k \textbf{R}} B_{k,\alpha,n} = \\
    &\sum_{\textbf{R},\tilde{\textbf{R}},\alpha,n} c_{\textbf{R},\alpha} w_{\textbf{R}^{\prime},\alpha',n} \epsilon_{\tilde{\textbf{R}},\alpha} \frac{1}{\sqrt{2\pi}} \sum_k e^{-i k (\textbf{R}-\tilde{\textbf{R}})} B_{k,\alpha,n} = \\
    &\sum_{\textbf{R}} \epsilon_{\textbf{R}-\textbf{R'},\alpha} c_{\textbf{R},\alpha}
\end{flalign}
where we have used Eq. (\ref{S:WanInReal}) and Eq. (\ref{S:EigenvalueEq}) for the first equality, Eq. (\ref{S:FourierSeriesEnergy}) for the second and Eq. (\ref{S:WanInReal}) and Eq. (\ref{S:WannierOrtho}) for the third. Plugging Eq. (S12) into Eq. (\ref{S:DNSEinWan}) directly gives Eq. (4) of the main text.

\section{Rice-Mele model}
In this section, we give further details about the Rice-Mele model \cite{Rice1982} used in the main text. The Rice-Mele model has two sites per unit cell and is the Thouless pump model with the smallest number of sites per unit cell. We schematically illustrate the model in Fig. S1a, where the intra-unit cell hopping is described via $J_\text{intra}(t) = -J - \delta \cos(\Omega t)$ and the inter-unit cell hopping via $J_\text{inter}(t) = -J + \delta \cos(\Omega t)$. The strength of the staggered on-site potential is given by $\Delta(t) = \Delta \sin(\Omega t)$. This is equivalent to Eq. (7) in the main text. The band structure of this model ($J=1$, $\delta=0.5$, $\Delta=1$) shows two bands (see Fig. S1b) with Chern number +1 and -1 for the lower and upper band, respectively. A simple way to calculate the Chern number is to plot the position of the Wannier centers for one pump cycle $t \in [0,2\pi/\Omega]$, as shown in Figs. S1c and d. 

\begin{figure}[htb]
\includegraphics{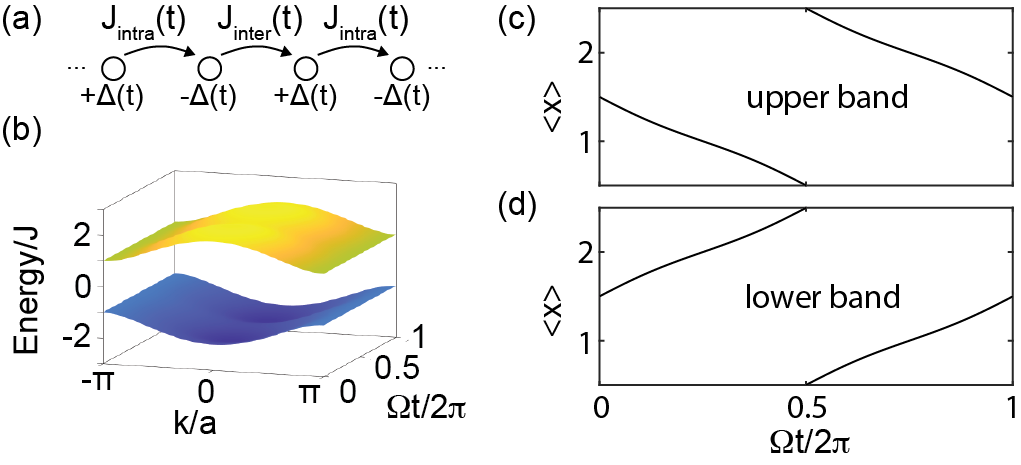}
\caption{Rice-Mele model. (a) Schematic illustration of the Rice-Mele model showing the inter-unit cell hopping ($J_\text{inter}(t)$), intra-unit cell hopping ($J_\text{intra}(t)$) and a staggered on-site potential $\Delta(t)$. (b) Band structure of the Rice-Mele model showing two bands. (c,d) Center of mass position of the instantaneous Wannier functions for the upper (c) and lower band (d) as a function of the pumping parameter $\Omega t$.}
\end{figure}

\section{Calculation of the solitons}

The discrete nonlinear Schr\"odinger equation is non-integrable (with only two conserved quantities). It has no known analytic soliton solutions, but they can be found numerically. We use two methods to calculate solitons: (1) Self-consistent iteration and (2) Newton's method. While the self-consistent algorithm works best for on-site solitons at high power, which are strongly localized, it will often not converge successfully to inter-site solitons and/or solitons at low power. In the main part, we use the self-consistent algorithm therefore to calculate the solitons of Figs. 1 and 4. In order to find low-power and/or inter-site solitons (as shown in Figs. 2,3 and Fig. S2) we use Newton's method. As convergence criterion for all calculated solitons, we use the following: $\sum_n |\psi^{\text{soliton}}_n-\psi^{\text{ev}}_n| \leq 10^{-25}$, where $\psi^{\text{soliton}}_n$ is the wavefunction of the calculated soliton after our final iteration step (be it self-consistent or Newton's method) and $\psi^{\text{ev}}_n$ is the eigenvector of the nonlinear Hamiltonian that has the largest overlap. Importantly, the convergence of both methods relies on suitable starting guesses. In the following we briefly describe both methods and suitable initial wavefunctions.

\subsection*{Self-consistency algorithm}

The self-consistent algorithm converges exponentially using the following iterative steps: 
\begin{enumerate}
  \setlength{\itemsep}{1pt}
  \setlength{\parskip}{0pt}
  \setlength{\parsep}{0pt}

  \item Choose an initial guess for the wavefunction, $\psi^{\text{(i)}}$. Typical initial guesses with a high success rate are single-site excitation or the Wannier function of the band from which the soliton is intended to bifurcate.
  \item Calculate the full Hamiltonian, which includes the linear Hamiltonian ($H_{\text{lin}}$) and the nonlinearly-induced potential: $H = H_{\text{lin}} - g |\psi^{\text{(i)}}|^2 $
  \item Calculate the eigenvectors of $H$ and choose the eigenvector with the largest overlap with $\psi^{\text{(i)}}$, updating the initial guess to be this new eigenstate.
  \item Iterate steps 2 and 3 until the desired convergence is achieved.
\end{enumerate}

\subsection*{Newton's method}
We use Newton's method to find low-power and inter-site solitons. To be specific, we solve the set of $N$ (with $N$ being the number of sites) equations describing the stationary discrete nonlinear Schr\"odinger equation for a given degree of nonlinearity $g$, while simultaneously constraining the power of the wavefunction.
\begin{flalign*}
    0 =& \sum_m H_{nm} \psi_m - g |\psi_n|^2 \psi_n - \lambda \psi_n  \qquad \forall n \in \{1,2,\dots,N\} \\
    0 =& \sum_n |\psi_n|^2 -1
\end{flalign*}
Here, $H_{nm}$ is the tight-binding real space Hamiltonian and the algorithm solves for the amplitudes of the wavefunction, $\psi_n$, and the nonlinear eigenvalue $\lambda$. A similar set of equations (that follows straightforwardly form Eq. (6)) is used when solving for the soliton in Wannier space. We use Mathematica's FindRoot \cite{Wolfram12} to solve the set of $N+1$ equations. Especially for low power, successful convergence critically depends on the initial guess. 

\begin{figure*}[t!]
\includegraphics{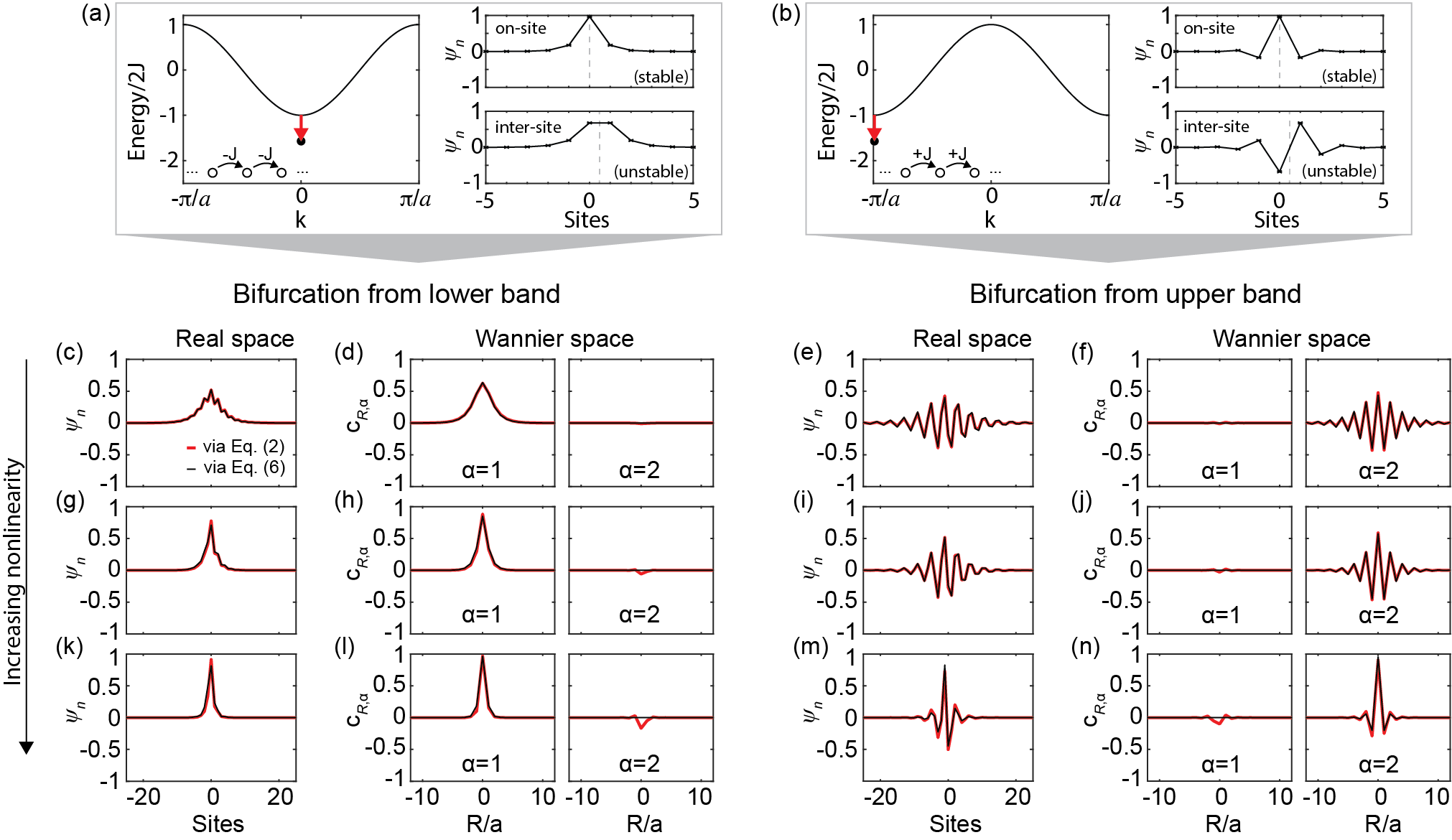}
\caption{Comparison between soliton calculation for increasing power. (a) Left panel: Band structure for a simple 1D lattice ($J>0$). Right panels: Illustration of a (stable) on-site and an (unstable) inter-site soliton. This band structure corresponds to the lower band in the Rice-Mele model. (b) Same as (a), but for a simple 1D lattice with positive nearest-neighbor hoppings, corresponding to the upper band in the Rice-Mele model. (c-n) Expansion coefficients of the solitons calculated in real space via Eq. (2) (shown in red) and in Wannier space via Eq. (6) (shown in black). The solitons are shown for increasing nonlinearity using $g/J=1$, $g/J=1.4$ and $g/J=2.5$ for subfigures (c-f), (g-j) and (k-n), respectively. Subfigures (c,d,g,h,k,l) are for solitons that bifurcate from the lower band. Subfigures (e,f,i,j,m,n) are for solitons that bifurcate from the upper band. $\alpha=1$ ($\alpha=2$) is the band index and refers to the lower (upper) band, respectively. Figs. S2c,d are identical to Figs. 2b,c.}
\end{figure*}

The solitons shown in Fig. 2 are first solved for in Wannier space using the following starting guess derived from the continuum approximation:
\begin{equation}
    c^{(0)}_{R,\alpha,n} = -\text{sign}[\epsilon_{R_1,\alpha}]^n \sqrt{\frac{ 2 \tilde{E} |\epsilon_{R_1,\alpha}|}{g W} } 
    \,\,\, \text{sech}(\sqrt{ \tilde{E}} \, (n - n_0))
\end{equation}
where $\epsilon_{R_1,\alpha}$ is the first Fourier coefficient of the band structure for band $\alpha$, describing the nearest-neighbor hoppings, $n_0$ is the center of the soliton, which can be chosen on-site or inter-site, $W \equiv W_{R,R,R,R}^{\alpha,\alpha,\alpha,\alpha}$ is the overlap integral as defined in Eq. (5), and $\tilde{E} = \left( \frac{P g W}{4 |\epsilon_{R_1,\alpha}|} \right)^2 $ is the energy difference between the (linear) band minimum and the energy of a continuum approximated soliton. The solitons found in Wannier space (shown in black in Fig. 2 and Fig. S2) are then transformed into real space and subsequently used as seeds to calculate the solitons in real space (shown in red in Fig. 2 and Fig. S2). When calculating solitons for a full pumping cycle (as shown in Fig. 3), we use the soliton of each time step as the starting guess for the next time step.


\section{Additional comparison between solitons calculated in real space and Wannier space}

In this section, we report additional calculations that compare the solitons calculated via Eq. (2) and Eq. (6) for the upper band of the Rice-Mele model and increasing nonlinearity (for identical model parameters as in the main text). While both inter- and on-site centered solitons exist, we only show the latter. The left (right) part of Fig. S2 shows the calculation of solitons that bifurcate from the lower and upper band. Figs. S2c,d are identical to Figs. 2b,c.

The main difference between solitons that bifurcate from the lower band and the upper band lies in the phase profile of the soliton. For focusing nonlinearity ($g>0$) solitons bifurcate from the bottom of the band, which is at $k/a=0$ ($k/a=\pm \pi)$ for the lower (upper) band in the given Rice-Mele model. This is the analogue of the bands in a simple 1D lattice, with negative (see Fig. S2a) and positive (see Fig. S2b) hoppings. As illustrated in Figs. S2f,j,n, the solitons that bifurcate from the upper band have a staggered phase profile in Wannier space. 

With increasing power the approximations of Eq. (6) compared to Eq. (2) become less accurate due to occupation of other bands (see Fig. S2l and Fig. S2n). Nevertheless, the general features of the soliton are still consistent with the low-power regime. Indeed, we expect quantized nonlinear Thouless pumping to be observable until a nonlinear bifurcation splits the path of the soliton (see also \cite{Juergensen2021}) or the solitons become unstable.


\bibliography{bibliography}